\documentclass[a4paper,10pt,twoside]{cpc-hepnp}

\usepackage{multicol}
\usepackage{graphicx}
\usepackage{booktabs}
\usepackage{amssymb,bm,mathrsfs,bbm,amscd}
\usepackage[tbtags]{amsmath}
\usepackage{lastpage}

\begin{document}

\fancyhead[c]{\small Chinese Physics C~~~Vol. XX, No. X (201X)
XXXXXX} \fancyfoot[C]{\small XXXXXX-\thepage}

\footnotetext[0]{Submitted to 'Chinese Physics C'}

\title{A Curve-based Material Recognition Method in MeV Dual-energy X-ray Imaging System\thanks{Supported by National Natural Science
Foundation of China (11235007 \& 10905030)}}

\author{%
      Chen Zhi-qiang$^{1,2}$%
\quad Zhao Tiao$^{1,2}$%
\quad LI Liang$^{1,2;1)}$\email{lliang@mail.tsinghua.edu.cn}
}
\maketitle

\address{%
$^1$ Department of Engineering Physics, Tsinghua University, Beijing 100084, China\\
$^2$ Key Laboratory of Particle \& Radiation Imaging (Tsinghua University), Ministry of Education, Beijing 100084, China\\
}

\begin{abstract}
High energy dual-energy X-ray Digital Radiography(DR) imaging is mainly used in material recognition of the cargo inspection. We introduce the development history and the principle of the technology and describe the data process flow of our system. The system corrects original data to get the dual-energy transparence image. Material categories of all points in the image are identified by the classification curve which is related to the X-ray energy spectrum. For the calibration of classification curve, our strategy involves a basic curve calibration and a real-time correction devoted to enhance the classification accuracy. Image segmentation and denoising methods are applied to smooth the image. The image contains more information after colorization. Some results show that our methods achieve the desired effect.
\end{abstract}



\footnotetext[0]{\hspace*{-3mm}\raisebox{0.3ex}{$\scriptstyle\copyright$}2013
Chinese Physical Society and the Institute of High Energy Physics
of the Chinese Academy of Sciences and the Institute
of Modern Physics of the Chinese Academy of Sciences and IOP Publishing Ltd}%

\begin{multicols}{2}

\section{Introduction}

The X-ray imaging technique has become one of the most important tools in customs inspection. At present, there are mainly two X-ray imaging modalities: radiography and computed tomography (CT). Although CT can provide 3-D structures and accurate attenuation map of the cargo, its complexity and high price limited its application \cite{lab1,lab2,lab3}. The X-ray radiography, including single-energy and dual-energy, is still the mainstream technology. The development of the X-ray radiography undergoes three stages: X-ray film photography, Computed Radiography(CR) and DR. The single-energy X-ray DR image merely gives the cumulative density information of the irradiated objects in one direction. It is used in preliminary medical diagnosis and simple security inspection. Since the single energy X-ray DR provides limited information, the dual-energy method was developed. Low energy dual-energy X-ray DR imaging has been widely used in current security inspection equipment which can detect and distinguish the contraband by determining material atomic number Z. The X-ray’s energy here is usually lower than 1MeV. This technology is inapplicable in high Z material recognition or cargo inspection as the energy of the X-ray which can penetrate the object in these situations needs to be a few MeV.

The British company Cambridge Imaging first proposed the idea of high energy dual-energy X-ray imaging. There was some disputes about the validity of the high energy dual-energy method in material recognition. The Russian Efremov research institute proved the feasibility of this method with their experimental prototype\cite{lab4,lab5,lab6,lab7,lab8}. The Germany company Heimann and the American company EG\&G applied the X-ray hardening technology to this field and proposed the filter method. The department of engineering physics of Tsinghua University and its cooperative enterprise Nuctech established an experimental platform and made some achievements on material recognition and related studies\cite{lab9,lab10,lab11,lab12,lab13,lab14,lab15,lab16,lab17}.

In this paper, we present a simple model of high energy dual-energy X-ray DR imaging system and focus on concrete methods in imaging and material recognition process. In section 2, we introduce our system model and the technology principle and elaborate the methods we use. We give some experiment results in section 3 and make a conclusion and discussion in section 4.

\section{Method}

\subsection{The principle of MeV dual-energy X-ray imaging in material recognition}
Three main ways of interaction of photon with matter are photoelectric effect, Compton scatter effect and electron pair effect. They respectively dominate the low($<1\text{MeV}$), middle($1\sim3\text{MeV}$) and high($>3\text{MeV}$) energy range. The corresponding attenuation coefficients $\mu$ have different dependences with material atomic number $Z$. Actually, we can give
\begin{eqnarray}
\label{eq1}
& \mu_{\text{P}} \propto Z^4 & \nonumber\\[1mm]
& \mu_{\text{CS}} \propto 1 &\\[1mm]
& \mu_{\text{EP}} \propto Z. & \nonumber
\end{eqnarray}
P, CS and EP are the abbreviations of three effects. Consider an X-ray source whose energy spectrum is $N(E)$ and the highest energy is $E_m$, a single substance with $Z$ and $\mu(E,Z)$, the transparence $T$ is
\begin{equation}
\label{eq2}
T = \frac{I}{I_0} = \frac{\int_0^{E_m} N(E)e^{-\mu(E,Z)}\,dE}{\int_0^{E_m} N(E)\,dE}.
\end{equation}
In dual-energy situation, the highest energy of two X-ray sources are $E_1$ and $E_2$. $R$ is defined as the logarithmic ratio of $T$.
\begin{equation}
\label{eq3}
R = \frac{\ln{T_1}}{\ln{T_2}} = \frac{\bar{\mu}(E_1,t,Z)}{\bar{\mu}(E_2,t,Z)}.
\end{equation}
$R$ is the ratio of equivalent $\mu$. When the energy spectrum is a single line, Eq.~(\ref{eq2}) and Eq.~(\ref{eq3}) can be simplified. Suppose $E_1$ in the low energy range and $E_2$ in the middle energy range, $R$ can be written as
\begin{eqnarray}
\label{eq4}
R  = \frac{\ln{T_1}}{\ln{T_2}} = \frac{\ln{e^{-\mu(E_1,t,Z)}}}{\ln{e^{-\mu(E_2,t,Z)}}} = \frac{\mu(E_1,t,Z)}{\mu(E_2,t,Z)}\nonumber \\
\approx  \frac{\mu_{\text{P}}(E_1,t,Z)}{\mu_{\text{CS}}(E_2,t,Z)} \approx c(E_1,E_2)Z^4.
\end{eqnarray}
The function $c$ is just concerned with $E_1$ and $E_2$. $R$ is greatly dependent on $Z$. Because $R$ is a distinct indication of $Z$ and it is calculated with the intensity value easily obtained from X-ray image, low energy dual-energy X-ray DR imaging technology has been widely used in small security inspection devices. The ‘low energy dual-energy’ means that E1 and E2 are usually lower than 1MeV.

When $Z$ is high and the irradiated objects are thick, the low energy X-ray imaging becomes useless. We change $E_1$ to the high energy range and keep $E_2$ in the middle energy range, then $R$ is
\begin{equation}
\label{eq5}
R = \frac{\mu(E_1,t,Z)}{\mu(E_2,t,Z)} \approx  \frac{\mu_{\text{EP}}(E_1,t,Z)}{\mu_{\text{CS}}(E_2,t,Z)} \approx c(E_1,E_2)Z.
\end{equation}
$R$ is dependent on $Z$ to a certain extent and also can be used to classify material. This ideally conclusion without consideration of the subordinated interaction of photon with matter is based on the assumption of the single-line X-ray energy spectrum. In fact, the accelerator as a high energy X-ray source generates the X-ray with a broad energy spectrum. Most of photons distribute in the middle energy range where $\mu$ and $Z$ have no correlations. The effectiveness of $R$ in material recognition is not obvious. It was found that although $R$ changes when thickness $t$ changes, $R$ is still dependent on $Z$\cite{lab4}. Here $E_1$ and $E_2$ are usually higher than 3MeV.

\subsection{A MeV dual-energy System}
We use an imaginary model (See Fig.~\ref{fig1}) including an accelerator which emits a vertical fan shaped X-ray beam, a scanning track which is perpendicular to the main beam direction, an L shape detector, and a data processing unit consisted of four parts. The unit first gets and corrects the original dual-energy X-ray images to obtain the dual-energy transparence images. By using the dual-energy transparence images and the classification curve, the material information image is formed. The smoothing process is to improve the image quality. Final part of the unit is the colorization of the gray image. In next four small sections, we focus on these four parts of data processing unit and mainly concentrate on the calibration of the classification curve. In section 3, we can see that our method enhances the classification accuracy and gives a satisfactory visual result.

\begin{center}
\includegraphics[bb= 0 0 8cm 4cm, trim=0 17pt 0 12pt,clip]{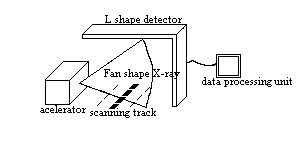}
\figcaption{ An imaginary model of the cargo inspection system } \label{fig1}
\end{center}

\subsection{Data acquisition and pre-processing}
Our model is ‘imaginary’ because the accelerator as an X-ray source produces the dual-energy X-ray simultaneously. Accordingly, the detector is able to distinguish high and low energy X-rays and form the dual-energy X-ray images separately. Other aspects of our model are basically same with reality. There is an angular distribution in the fan shaped X-ray of which main beam located near the middle of the fan has the maximum intensity decreasing towards both sides. It makes different vertical position of the X-ray image have different X-ray intensity and energy spectrum. The accelerator state fluctuation in scanning process makes different lateral position of the X-ray image have different X-ray intensity and energy spectrum. The detector background and response inconsistencies also exist. In data processing unit, the first step is to correct the original data and get the dual-energy transparence image. We give Eq.~(\ref{eq6}) based on Eq.~(\ref{eq2}).
\begin{equation}
\label{eq6}
T(x,y) = \frac{I(x,y)-I_{\text{BK}}(x,y)}{I_0(x,y)}\cdot LD(x).
\end{equation}
The coordinate $x$ and $y$ represents the lateral and vertical position. $I$ is the original dual-energy image. $I_0$ is the air image. $I_{\text{BK}}$ is the detector background image. The correction factor $LD$ obtained by monitoring the accelerator state fluctuation in scanning process is a function of the lateral position x. In Eq.~(\ref{eq6}), the division by $I_0$ corrects the intensity angular distribution and detector response inconsistencies. Rest corrections are also done by Eq.~(\ref{eq6}). After the point-by-point correction and simple denosing, we can get the dual-energy transparence image $T$.

To take advantages of two transparence images, we use a point-wise weighted fusion of them. In the thick places of the irradiated object, the high energy transparence image give more information than the low energy transparence image. The conclusion is opposite in the thin places. The gray value representing one point’s thickness determines the weight. Using the dual-energy transparence image and the classification curve which we elaborate next, we can get the material information image. The colorization assembling the fused transparence image and the material information image gives a final result.

\subsection{Curve-based material recognition method}
\begin{figure}[ht]
\centering
\includegraphics[bb=0 0.4cm 15cm 10.6cm]{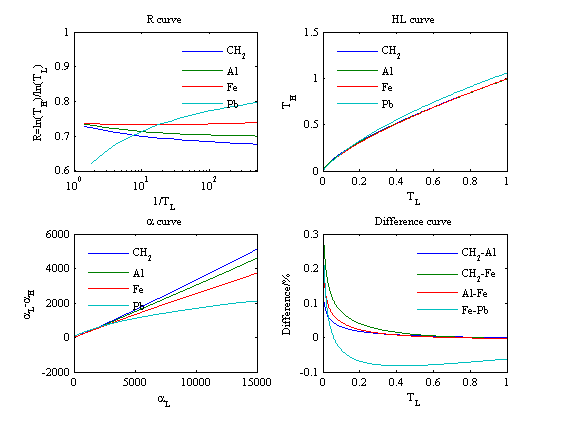}
\caption{ Four kinds of classification curves. The kind derived from R curve has the best visual separability }\label{fig2}
\end{figure}

There are four kinds of classification curves. They are essentially same and the difference of them can be completely eliminated by coordinate transformation. Among them, one derived from R curve has the best visual separability (See Fig.~\ref{fig2}).
\end{multicols}
\begin{multicols}{2}
The longitudinal coordinate of R curve is $R$ defined in Eq.~(\ref{eq3}) and the horizontal coordinate is the inverse of $T_1$ or $T_2$. Different materials have different R curves. They are arranged in the order of increasing $Z$ when they are in the same image. The classification curve here has several single R curves related to the same nufig2mber of typical objects. We assume that there are four typical objects including Pb, Fe, Al and CH2(or C) as we can see in Fig.~\ref{fig2}.

$(x,y)$ is one point of the dual-energy transparence image. It has two transparence values $T_1(x,y)$ and $T_2(x,y)$. They give a point $(R,1/T2)$ on the classification curve C. Because R curves of different materials are arranged in the order of increasing $Z$, $Z$ of this point is obtained by interpolating between two adjacent R curves in C. Repeat this procedure on every point of dual-energy transparence image, the material information image $Z(x,y)$ is formed.

Eq.~(\ref{eq2}) and Eq.~(\ref{eq3}) show that R curve correlates to the energy spectrum. We already know that each point of the dual-energy transparence image has different X-ray intensity and energy spectrum because of the angular distribution and the accelerator state fluctuation. This fact causes a problem that different points of the dual-energy transparence image need different classification curves. However，it is impossible to calibrate all of classification curves.

The calibration strategy we employ takes two steps. First we get the basic classification curve before the system scans cargo or after the system state significantly changes. This step requires scanning several typical objects. In one scanning process, the system only scans one object with different mass thickness. Like the common scan, we get two transparence image $T_1$, $T_2$ and each point $(x,y)$ belongs to a R curve $R_{xy}$. Assuming a steady system state in the process, we ignore the differences of energy spectrum in different lateral positions and just use several vertical points to represent the whole angular distribution. So we have
\begin{equation}
\label{eq7}
R_{xy}=R_y,y=y_1,y_2,...,y_n.
\end{equation}
The data is
\begin{eqnarray}
\label{eq8}
& data_y=\left\{R_y(T_1(x_i,y),T_2(x_i,y)),i=1,2,...,m \right\} &\nonumber \\
& y=y_1,y_2,...,y_n. &
\end{eqnarray}
$m$ is the number of different mass thickness and $n$ is the number of vertical points. The R curve is fitted with these data and the classification curve is formed with the R curves of several typical objects. Here, we use Pb, Fe, Al, C.
\begin{eqnarray}
\label{eq9}
& \left\{R_y^Z|R_y^Z=fit(data_y^Z),y=y_1,y_2,...,y_n\right\} &\nonumber \\
& C_{y_i}=\left\{R_{y_i,Z}|Z=\mathrm {Pb,Fe,Al,C}\right\},i=1,2,...,n. &
\end{eqnarray}

In the cargo scan, system monitors the state variation to complete the real-time calibration of the classification curve. A small device consisted of the typical objects with single thickness is set in a certain vertical position Y and scanned synchronously with the cargo. The data is
\begin{eqnarray}
\label{eq10}
& T_{1r}^Z(x_i,Y),T_{2r}^Z(x_i,Y) &\nonumber \\
& i=1,2,...,nx,Z=\mathrm{Pb,Fe,Al,C}. &
\end{eqnarray}
$nx$ is the number of horizontal pixels. In the first step, this data is also saved as
\begin{equation}
\label{eq11}
T_{1b}^Z(x_i,Y),T_{2b}^Z(x_i,Y).
\end{equation}
They are the average of $T$ on the horizontal direction as we ignore the lateral difference. Then the revised classification curve will be
\begin{eqnarray}
\label{eq12}
&C_{x_jy_i} = \left\{R_{x_jy_i}^Z|R_{x_jy_i}^Z=R_{y_i}^Z \ast F(T_{1r}^Z,T_{2r}^Z,T_{1b}^Z,T_{2b}^Z;x_j,Y),\nonumber \right. &\\
& \left.Z=\mathrm{Pb,Fe,Al,C} \vphantom{R_{x_j}^Z}\right\} &\\
& i=1,2,...,n,j=1,2...,nx. &\nonumber
\end{eqnarray}
The function $F$ use the real-time calibrating data to estimate the difference between $R_{xy}^Z$ and $R_y^Z$. The correction method $\ast$ in Eq.~(\ref{eq7}) relies on how function $F$ is calculated. Considering the statistical straggling of $T(x_j,Y)$, we choose a segment of $x_j$ and use the average of the monitoring data to correct the classification curve.

\subsection{Smooth of material information image}
The detector signal noise inevitably exists and it is even amplified in the data processing. It has impact on material recognition accuracy and makes the material information image rough. The visual effect of final result after colorization is not good enough. The quality promotion of the material information image is necessary.

First step of our smooth process is the segmentation of the fused transparence image which is obtained previously. The image is segmented into numbers of regions which keep the continuity of the irradiated objects interior as much as possible and discriminate different irradiated objects as clear as possible. The average of all the $Z$ values in a corresponding region in the material information image is assigned to all points in this region.

General image segmentation algorithm like single-pass spilt-merge algorithm, or data clustering algorithm like the Leader algorithm can be used here with some adjustment\cite{lab18}. The irradiated objects may be mixed and disorderly. So the segmentation or cluster result may have too many small areas with only several pixels. This over-segmentation can be solved by merging the small area into the nearest large area. The ‘nearest’ means not only the distance but also the similarity between them.

Using the average of $Z$ values in one segmentation region to replace all points in this region brings some loss of original information. The denoising approaches can give a better material information image and the majority of the image remains the same. It is a chanllenge to find a better method which can smooth the image while maximizing the retention of the original information.

\subsection{Colorization}
In the colorization, different colors represent different materials. We use three color spaces including RGB, HLS and YUV. If all three values in a color space are known, a color is determined.
We divide the range of $Z$ into several parts as follow,
\begin{equation}
\label{eq13}
Z<Z_1;Z_1\leq Z_2;...;Z_{p-1}\leq Z_p;Z_p\leq Z.
\end{equation}
There are $p+1$ hues $H_1$, $H_2$, $...$, $H_p$, $H_{p+1}$. When the $Z$ value of one point in the material information image falls into the $j$th part, the $H$ value in HLS color space equals to $H_j$.

The sensitivity of the human eye to different color is different. There are red, green and blue with the same $L$ value in HLS color space. The brightness felt by eyes is different. Green is the brightest and red is brighter than blue. If $L$ value equals to the gray value, the points with same gray value will give different brightness when we look at the final result. It is inappropriate.

In YUV color space, colors with same $Y$ value give closest brightness feeling. Let $Y$ value equal to the gray value of the fused transparence image. YUV color space and HLS color space can be converted to each other. So we have
\begin{equation}
\label{eq14}
Y  = f(H,L,S).
\end{equation}
As $Y$, $H$, $f$ is known and $S$ is given, $L$ is the solution of Eq.~(\ref{eq14}). Then all three values in HLS color space are already set, a color is determined. Repeat this procedure on every point of the material information image to get final result.

The table of hue and the saturation value $S$ is changeable and directly related to the image visual effect. The mapping relationship between $Y$ and the gray value can be optimized by some transformation like logarithm stretching.

\section{Experiment result}
The data is provided by a 6/3MeV X-ray DR imaging cargo inspection system which applies our basic design of data processing unit. The accelerator alternatively emits the high and low energy X-ray. The emission frequencies are both 40Hz. We use a single column of CWO detector. Scanning speed is 0.2m/s.
The calibration of the basic curve is based on the scanning data of material C, Al and Fe. The system scans single substance with 15 different mass thickness to form one R curve. There are two different system states. In Fig.~\ref{fig3}, the dotted curve represents the basic curve in state 1 and the dashed curve represents the basic curve in state 2. The solid curve represents the revised classification curve in state 1 using the difference between the real-time calibrating data of two states. We think that the dashed curve is the ‘true’ classification curve in state 2 and the solid curve is the estimation of the ‘true’ one. They are close and show the effectiveness and rationality of our calibration strategy.
\begin{center}
\includegraphics[bb=0 0.5 8cm 6cm]{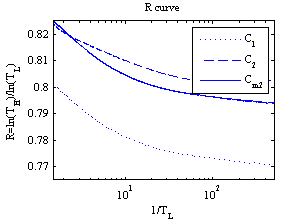}
\figcaption{\label{fig3}   The 'true' curves in two states are $\text{C}_1$and$\text{C}_2$. The revised R curve using the real-time calibration data to estimate the 'true' curve in state 2 is $\text{C}_{\text{m}2}$}
\end{center}

We arrange eight objects in the order of Pb, Fe, Al, C, Al, C, Pb, Fe and divide them into two groups according to the size. Their mass thickness are all 40g/$\text{cm}^2$. The scanning of these eight objects is in state 2. Suppose we do not know the basic curve in state 2(dashed curve in Fig.~\ref{fig3}). In Fig.~\ref{fig4}, the image a) on the left is the final result without the use of the real-time calibrating data(using dotted curve in Fig.~\ref{fig3}), the image b) in the middle is the final result with the use of the real-time calibrating data(using solid curve in Fig.~\ref{fig3}). The image c) on the right is the color table. It is the system colorization settings which sets the hues of four typical objects C, Al, Fe and Pb are orange, green, blue and purple. The comparison clearly shows the effectiveness of the real-time calibration and also matches the comparison of the curves in Fig.~\ref{fig3}. Note that there is no R curve of Pb in the classification curve, the colorization of Pb is deviated from the righteous color. Besides Pb, other materials’ results show that our calibration strategy enhances the classification accuracy.
\end{multicols}
\begin{center}
\includegraphics[bb=0 0cm 15cm 3cm]{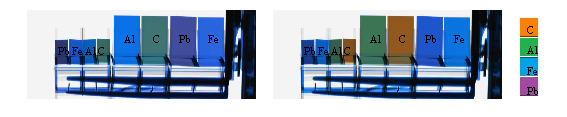}
\figcaption{\label{fig4}   Comparison of the result with and without the real-time calibration. a), without. b), with. c), color table. }
\end{center}
\begin{multicols}{2}
In Fig.~\ref{fig5}, the comparison is the final color image with and without the smooth process. The larger orange object in the image a) looks not uniform although it should be the same color. We can see that the smoothing improves the image quality. But the effect is obsolete because of the monotony and uniformity of the irradiated objects.
\end{multicols}
\begin{center}
\includegraphics[bb=0 0cm 15cm 3cm]{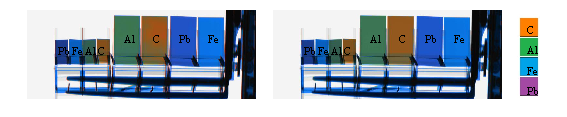}
\figcaption{\label{fig5}   Comparison of the result with and without smoothing.  a), without. b), with. c), color table. }
\end{center}
\begin{multicols}{2}
In Fig.~\ref{fig6}, we give a cargo inspection result. The emission frequencies of the dual-energy X-ray are both 33Hz. The scanning speed is 0.2m/s. The irradiated objects from left to right are cigarette, salt, sugar, coffee, buckets of water, and concrete. According to the continuity of the irradiated objects, we can assure that the spots and stripes in the object regions of the image on the top are noise and need to be removed. The smooth process eliminates the noise spots and nonuniformity and significantly promotes the image quality. In red circled region of the image on the top, the bottom margin of the bucket is overwhelmed by the noise and almost disappeared. After the smooth process, the margin is recovered. Our smooth method may strengthens the details of simple object regions.
In the cigarette region, the thickness of cigarette is small. When the irradiated object is thin, the calibration curve separability is worse. With the data fluctuation, the final color image will be full of stripes and spots as we can see in the amplified region of the image on the top. The cigarette belongs to the orange category, so the blue pixels are the noise. The smooth process take the average of $Z$ in the cigarette region of the material information image. So the final color will be the middle color between orange and blue according to the color table. The color changes to green as we can see in the amplified region of the image on the bottom.
\end{multicols}
\begin{center}
\includegraphics[bb=0 0cm 15cm 10cm,trim=45pt 0pt 45pt 0pt,clip]{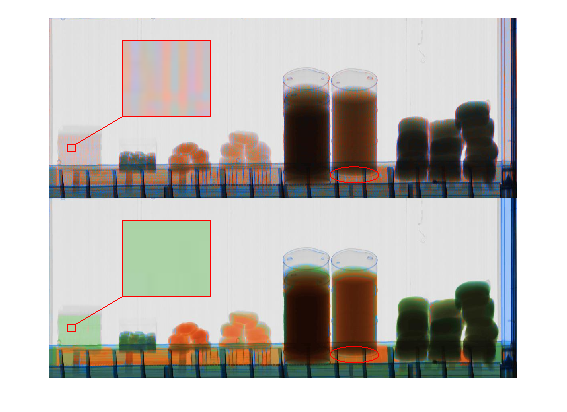}
\figcaption{\label{fig6}   Comparison of cargo inspection result with and without smoothing. Top, without. Bottom with. From left to right, the irradiated objects are cigarette, salt, sugar, coffee, buckets of water and concrete. The circled regions clearly show the smoothing effect.  }
\end{center}
\begin{multicols}{2}
\section{Discussion and conclusion}
We describe a simple MeV dual-energy X-ray DR imaging cargo inspection system with detailed description of the data processing unit. The preliminary treatment converts the original data into the images for subsequent processing needs. The calibration strategy of the classification curve enhance the classification performance. There are two points to note. First, the calibration of basic classification curve which is used as the basis needs a steady system state. Second, the real-time monitoring data has statistics fluctuation even exceeding the state fluctuation of the accelerator. Using the average of a piece of data to reflect the variation of this piece of data is better than using every single point to revise the curve. An improved way is to extract the variation tendency by overall consideration of the entire real-time monitoring data. The classification curve is revised by the variation tendency. The smoothing of the material information image is to enhance the image quality. Segmentation is the key part of our smooth process. Better segmentation method leads to better image quality. Color image can carry more information and give better visual effect. Colorization can be regulated in different application environment. This system design has a certain guiding significance to the engineering and further development of the dual-energy X-ray imaging technology.

\end{multicols}

\begin{multicols}{2}

\end{multicols}

\clearpage

\end{document}